\documentstyle[eqsecnum,aps,prc]{revtex}
\begin{document}
\draft
\title{Chaos modified wall formula damping of the surface motion
of a cavity
undergoing fissionlike shape evolutions}
\author{Santanu   Pal\thanks{Electronic address:santanu@veccal.ernet.in}
and Tapan Mukhopadhyay\thanks{Electronic address:tkm@veccal.ernet.in}}
\address{Variable  Energy  Cyclotron  Centre,  1/AF Bidhan Nagar,
Calcutta 700 064, India}
\date{\today }
\maketitle
\begin{abstract}
   The chaos weighted wall formula developed earlier for systems with
partially chaotic single particle motion is applied to large
amplitude collective motions similar to those in nuclear fission.
Considering an ideal gas in a cavity undergoing fissionlike shape
evolutions, the irreversible energy transfer to the gas is
dynamically calculated and compared with the prediction of the chaos
weighted wall formula. We
conclude that the chaos weighted wall formula
provides a fairly accurate description of one body dissipation
in dynamical systems similar to fissioning nuclei.
We also find a qualitative similarity between the phenomenological
friction in nuclear fission and the chaos weighted wall formula. This
provides further evidence for  one body nature of the dissipative force
acting in a fissioning nucleus.

\end{abstract}


\eject

\section{INTRODUCTION}

   It is now well established
   \cite{BBN,SIERK,NIX1,NIX2,NIX3,FROB1,FROB2,ABE,BW} that a dissipative
force operates in the
dynamics of a fissioning nucleus as it descends from the saddle point
to the scission. Since the fission process is expected to follow the mean
field dynamics when the excitation energy of the nucleus is not too high,
the physical processes that give rise to dissipation in fission
are usually identified as one body mechanisms.
The systematics of experimental mean kinetic energies of the
fission fragments were fairly well reproduced \cite{BBN,SIERK,BW}
using an
one body dissipative force in classical dynamical calculations.
However it has also been observed that the mean
kinetic energy of the fission fragments is not very sensitive to the details
of the dissipative forces. Though two body processes are less favoured
compared to one body processes in the long mean free path dominated
mean field regime,
both one and two body dissipations in classical
dynamical calculations have been found \cite{SIERK,BW} to describe
the systematics of experimental mean kinetic energies.
Nix and Sierk, on the other hand, suggested \cite{NIX1,NIX2,NIX3}
in their analyses of mean fragment kinetic energy data that the
effective dissipation is about four times weaker than that predicted by the
wall plus window formula of one body dissipation.

It has been pointed out \cite{FROB1,FROB2,ABE} in recent years that
prescission neutron and $\gamma$-multiplicities
are sensitive to the details of the
friction force in nuclear fission. In particular, it was shown \cite{ABE}
that one body dissipation is preferred to two body viscosity in order to
describe the neutron multiplicity data.
It was further observed \cite{FROB1} that a shape dependent
dissipative force is essential for simultaneous reproduction of experimental
values of prescission $\gamma$-multiplicities and fission probabilities.
Specifically, it has been observed \cite{FROB1,FROB2}
that the dissipative force in a compact configuration
of the fissioning compound nucleus should be much
smaller than that predicted by the wall formula of one body dissipation.

   The wall formula was developed by Blocki et al. \cite{BBN} in a
simple classical
picture of one body dissipation in which energy is transferred from
the nuclear surface motion to the nucleon motion as a result of frequent
collisions of the nucleons with the nuclear surface.
 According to the wall formula, the rate
of energy dissipation is given as

\begin{equation}
\dot{E}_{\text{WF}}(t)=\rho\bar{v}\int \dot{n}^2 d\sigma ,
\label{seqn1}
\end{equation}

\noindent
where $\dot{n}$  is  the normal component of the surface velocity at
the  surface element $d \sigma$ while the nuclear mass
density  and  the  average  nucleon  speed  inside  the   nucleus
are denoted by $\rho$ and $\bar{v}$
respectively.  The  wall  formula was also obtained from a formal
theory of one body dissipation based on classical linear response
techniques \cite{KR}.

      The wall formula was originally derived for idealized systems
employing a number of simplifying assumptions such as approximating the
nuclear surface by a rigid wall and considering only adiabatic collective
motions. The validity of these assumptions were scrutinised \cite{YAN,GD}
in the framework of random phase approximation (RPA) damping.
It was shown \cite{YAN,GD} that in the limiting situation where the
above assumptions are realized, RPA damping coincides with the wall
formula. These works further brought out the importance of including the
2p2h states for realistic applications.

We are however of the opinion that since the wall formula captures all the
essential ingredients of dissipation under mean field dynamics, it is still
possible to improve upon the dissipation rate given by Eq.(\ref{seqn1})
by examining its various assumptions. One such assumption of the wall
formula concerns the randomization of the particle motion.
It is assumed \cite{BBN} that successive collisions of
a  nucleon  with  the  one body potential give rise to a velocity
distribution which  is  completely  random.
In  other  words,  a
complete  mixing  in  the  classical  phase space of the particle
motion is required.
We have recently discussed \cite{PM} the implications of these assumptions
and obtained a modification of the wall formula by relaxing
the full randomization assumption
in order to make it applicable to systems in which the mixing  in
phase  space is partial. Considering  chaotic particle trajectories
which arise due to  irregularity in the shape of the one body potential
and which are responsible for irreversible energy transfer, we
modified the wall formula in Ref.\cite{PM} as,

\begin{equation}
\dot{E}_{\text{CWWF}}=\mu\rho\bar{v}\int \dot{n}^2 d \sigma ,
\label{seqn2}
\end{equation}

\noindent
where  $\mu$  is a measure of chaos (chaoticity) in the single particle
motion. The chaoticity $\mu$ is $1$ for  cavities  of  highly  irregular
shapes  in   which most of the
particles  move in a chaotic fashion,
and  one  recovers   the   original   wall   formula   given   by
Eq.(\ref{seqn1}) for such systems.
Eq.(\ref{seqn2})  becomes  useful  however for
systems in which the particle motion is partially chaotic  giving
rise to a chaoticity less than $1$. In fact, Eq.(\ref{seqn2}) gave
a quite successful description \cite{MP} of energy transferred to a gas from
the of its container oscillating with small amplitudes.
In what follows, we shall use
the    term    ''chaos   weighted   wall   formula''   (CWWF)   for
Eq.(\ref{seqn2}) in order to distinguish  it  from  the  original
wall formula (WF) given by Eq.(\ref{seqn1}).

      In the present work, we shall consider an ideal gas in a cavity
which will be subjected to a time dependent quadrupole deformation. We shall
calculate the energy transferred to the particles from the wall motion.
We  shall then compare this
transferred energy with that predicted by the chaos weighted wall
formula (Eq.(\ref{seqn2})). In this model,
the cavity will be progressivly deformed until the neck radius
vanishes.
The motion will be thus tailored to resemble
that of a fissioning nucleus as it descends from saddle to scission.
Our motivation in this work is twofold. First, we wish to test the
applicability of the chaos weighted wall formula for large amplitude
motions such as fission. Second, we would like to compare the chaos
weighted wall formula with the phenomenological reduced friction
suggested by Fr\"{o}brich et al.\cite{FROB2}. We shall find that the
chaos weighted wall formula can account for
the energy damping fairly accurately in
large amplitude collective motions in an ideal gas. We shall also
observe a qualiatative similarity between the chaos weighted
wall formula and the phenomenological reduced friction.

The dynamics of independent particles in time dependent cavities has
been extensively studied earlier by Blocki and his
coworkers \cite{BSS3,BBS1,BJS,BBS2,BSS1,BSS2,MSB,BBSS}. Considering
classical particles in vibrating cavities of
various shapes, a strong correlation between chaos in classical phase
space and the efficiency of energy transfer from collective to
intrinsic motion was numerically observed \cite{BSS3}. In particular, it
was demonstrated \cite{BSS3,BBS1} that while the energy transfer is much
smaller than the wall formula limit in a cavity undergoing quadrupole
vibration, it reaches the wall formula limit for higher multipole
vibrations.
Similar conclusions were also reached \cite{BBS2,BSS1,BSS2,MSB}
when the particle motion was treated quantum mechanically, though
the quantal energy transfers were found to be suppressed compared to
the classical ones. In the present work, we shall demonstrate that
the chaos weighted wall formula can describe the energy transferred
to classical particles when a cavity is strongly deformed starting
from spherical shape.

      The paper is organized as follows. We shall describe the
fissionlike cavity motion in the next section. A procedure
to obtain the irreversible energy transfer to the gas from the
cavity motion will be outlined in Sec.III.
Section IV will contain a comparison
between this irreversible energy transfer
and the chaos  weighted  wall  formula  prediction.
A summary of our main results will be presented in Sec.V where
we shall also draw our conclusions.

\section{A MODEL FOR FISSIONLIKE CAVITY MOTION}

We shall consider a gas  of  noninteracting particles in an axially
symmetric cavity which is undergoing a quadrupole deformation.
The initial velocity distribution for the particles will be assumed
to be uniform upto a certain maximum value $v$
similar to that of a completely degenerate Fermi gas.
The deformation of the surface will
be  proportional to the Legendre polynomials $P_2$ and the surface at
any instant $t$ will be defined as

\begin{equation}
R(\theta,t)=\frac {R_{0}}{\lambda (t)} \Big\{ 1+
\alpha(t)    P_2(\cos
\theta)\Big\} ,
\label{seqn3}
\end{equation}

\noindent
where

\begin{equation}
\alpha(t)= \alpha_0 - \alpha_2 \cos \omega t.
\label{seqn4}
\end{equation}

Here   $\alpha_0$ and  $\alpha_2$ are parameters which define the
amplitude of deformation and $\omega$ denotes its frequency while
$\lambda(t)$  is  a volume conserving
normalization  factor. According to Eq.(\ref{seqn4}), the rate of change of
deformation will be given as $\dot{\alpha}=\alpha_2 \omega\sin \omega t$.
The cavity
is thus initially at rest with a deformation $(\alpha_0 - \alpha_2)$
which subsequently changes  faster with time till the deformation
reaches $\alpha_0$ at $\omega t=\frac{\pi}{2}$. If we choose
$\alpha_0=2.0$ at which the neck radius becomes zero, the above
cavity motion can serve as a model for fission dynamics during its
transition from saddle to scission. By choosing a suitable value
for $\alpha_2$, the cavity shape at the saddle point can also be
defined.

We can now evaluate Eq.(\ref{seqn2}) for a quadrupole deformed cavity
and obtain

\begin{equation}
\Bigg(\frac{\dot{E}}{E_0}\Bigg)_{CWWF}
=\frac{15}{4}\alpha_2 \eta \omega \mu\big(\alpha(t)\big)
I\big(\alpha(t)\big) \sin^2\omega t ,
\label{seqn5}
\end{equation}

\noindent
where

\begin{equation}
I\big(\alpha(t)\big)= \frac{1}{\lambda^4 \big(\alpha(t)\big)}
\int_{-1}^{+1} F(z)dz ,
\label{seqn6}
\end{equation}

\noindent
and

\begin{equation}
F(z)=\frac{ [P_2(z)-\frac{1}{\lambda}(\frac{d\lambda}{d\alpha})
\big(1+\alpha P_2(z)\big)]^2 \big(1+\alpha P_2(z)\big)^3 }
{[\big(1+\alpha P_2(z)\big)^2 +\alpha^2 P^{\prime^2}_{2} (z)
          (1-z^2)]^{1/2}}.
\label{seqn7}
\end{equation}

In the above equations,
the adiabacity
index $\eta$ is defined \cite{BSS3} as the ratio of  the  maximum
speed  of  the surface tips to the maximum
particle speed $v$ and will be given by

\begin{equation}
\eta=\frac{\alpha_2 \omega R_0}{v},
\label{seqn8}
\end{equation}

\noindent
and  $E_0$ denotes the total
initial energy of the unperturbed gas. It may further be noted that
for  $\alpha \ll 1$, Eq.(\ref{seqn5}) reduces to

\begin{equation}
\Bigg(\frac{\dot{E}}{E_0}\Bigg)_{CWWF}
=\frac{3}{2}\alpha_2 \eta \omega \mu\big(\alpha(t)\big)
 \sin^2\omega t,
\label{seqn9}
\end{equation}

\noindent
which was used \cite{MP} earlier for small amplitude shape oscillations.
In the present work however we shall use Eq.(\ref{seqn5}) since we
shall be dealing here with large amplitude deformations. The essential
steps to obtain Eq.(\ref{seqn5}) can be found in the Appendix.
Subsequently the energy dissipated from the wall motion
after a certain interval $t$ will
be obtained as

\begin{equation}
\Bigg(\frac{E_{diss}(t)}{E_0}\Bigg)_{CWWF}
=\frac{15}{4}\alpha_2 \eta \int_0^t  \mu\big(\alpha(t')\big)
I\big(\alpha(t')\big) \sin^2\omega t'd(\omega t').
\label{seqn10}
\end{equation}

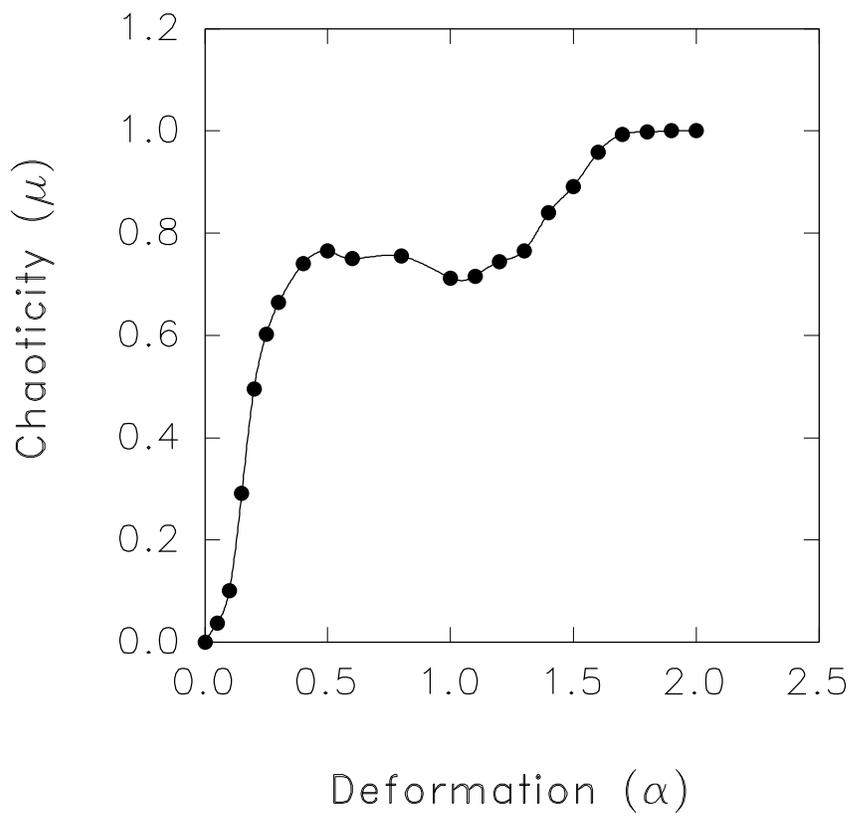
\begin{figure}
\caption
{Variation of the chaoticity with deformation for
quadrupole  shapes. Solid circles are calculated values
and the line is to guide the eye.}
\label{fig1}
\end{figure}

In order to calculate
the  dissipated energy using Eq.(\ref{seqn10}),  the  chaoticity
$\mu(\alpha(t))$   is
required for all deformations $\alpha(t)$ through which the
cavity  evolves with time.
The chaoticity for each deformation will be obtained \cite{PM} by
considering  particle  trajectories  in  a static cavity with the
same deformation  and  distinguishing  between  the  regular  and
chaotic  trajectories. Following the procedure outlined in
Ref.\cite{BBSS,MP},
the chaoticity $\mu$ will be determined by
 uniformly  sampling  of the  trajectories  which
originate  from  the  cavity  surface. Fig.\ \ref{fig1} shows the
calculated values of $\mu$ which  will be subsequently employed
to calculate the dissipated energy according to the chaos weighted
wall formula. It may also be noted that the original wall formula
dissipation will be obtained by simply putting $\mu=1$ in
Eq.(\ref{seqn10}).

\section{REVERSIBLE AND IRREVERSIBLE ENERGY TRANSFERS}

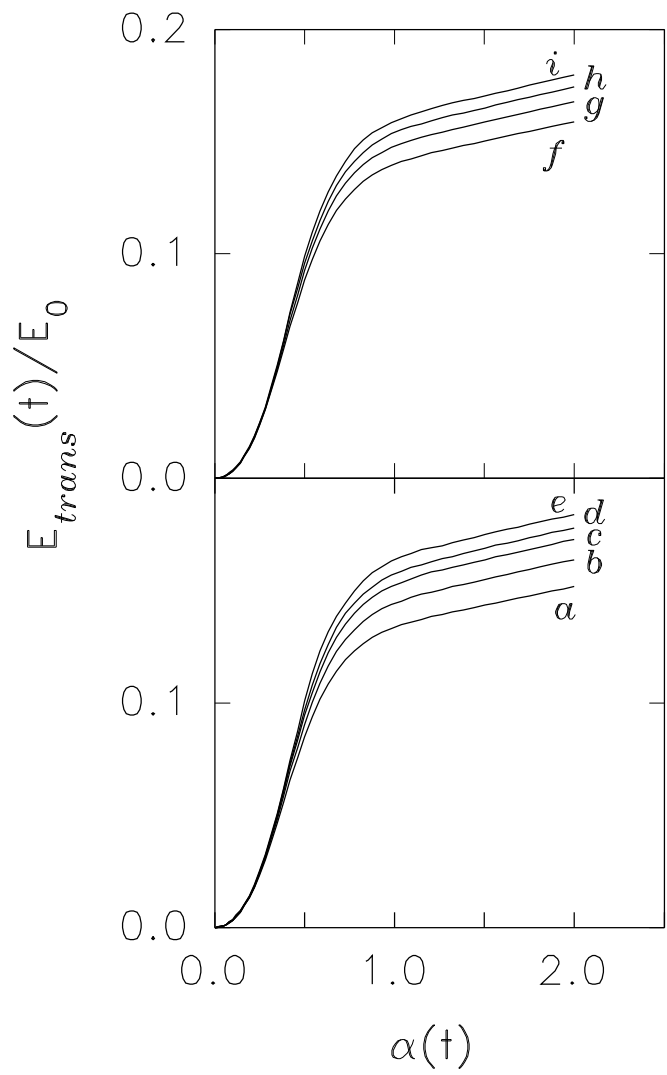
\begin{figure}
\caption
{Growth of relative energy transferred to the gas from the wall motion.
Curves $a, b, c, d$ and $e$ in the lower panel are for $\eta=0.02,
0.04, 0.06, 0.08$ and $0.10$ respectively while curves $f, g, h$
and $i$ in the upper panel are for $\eta = 0.03, 0.05, 0.07$ and $0.09$
respectively. The cavity motion is defined by
$\alpha_0 = \alpha_2 =2.0$.}
\label{fig2}
\end{figure}

In the dynamical calculation of
energy transferred to the particles in a
cavity from its surface motion, the classical
equation of motion of the particles is numerically solved.
To this end, the initial position
and the velocity vectors (lying within a Fermi sphere  of  radius
$v$ in velocity space) of a particle are chosen at random and the
trajectory   of   the   particle  is  follwed  in  time  allowing
reflections whenever it encounters the oscillating walls  of  the
cavity  \cite{BSS3}.  By  considering a gas consisting of a large
(typically  several  thousands)  number  of  such  noninteracting
particles in the cavity, the relative energy transferred to the
gas $E_{trans}(t)/E_0$ is obtained upto an interval given by
$\omega t=\frac{\pi}{2}$ at which the deformation becomes
maximum.  Fig.\ \ref{fig2}
shows the transferred energy calculated for different values of
$\eta$. The deformation parameters chosen are
$\alpha_0 = \alpha_2 =2.0$ which correspond to an initial spherical
shape and a maximum deformation of $\alpha=2.0$ at which the neck
radius is zero.

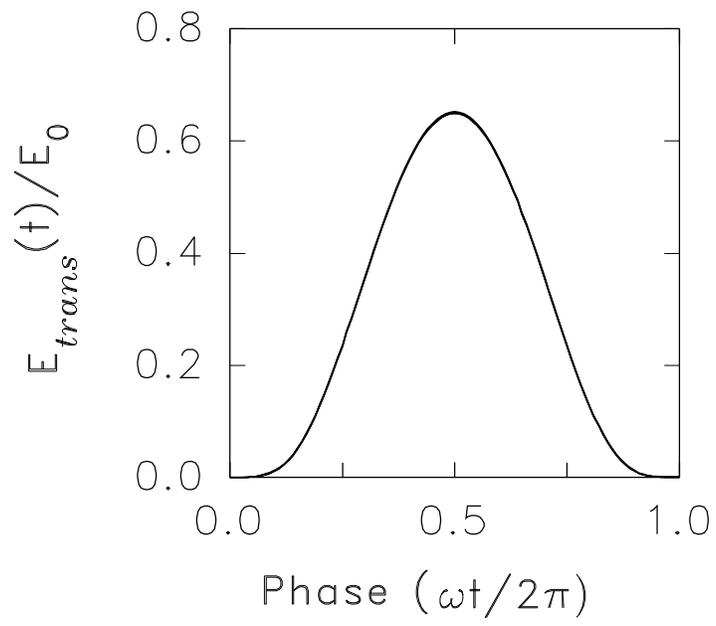
\begin{figure}
\caption
{Relative energy growth for a cavity undergoing spheroidal deformation
showing the complete reversibility of the energy transferred.
The plot represents calculated values for $\eta = 0.02, 0.04, 0.06, 0.08$
and $0.10$ which are all indistinguishable from each other. }
\label{fig3}
\end{figure}

      A part of the transferred energy $E_{trans}(t)$ is reversible
and is elastic in nature. This reversible part arises due to the
symmetries present in the single particle Hamiltonian of the system
and has been discussed by Blocki et al. \cite{BBN} in detail. For volume
conserving systems, the reversible energy depends only on the deformation
of the system and is independent of the speed of the deformation.
It is of interest to note here that this property is quite distinct \cite{BSS3}
from that of a real gas. An intrinsic randomization of
particle velocities is always assumed for an idea gas in a real container
giving rise to the ideal gas laws according to which the energy
of a gas does not change under adiabatic volume conserving deformations.
The reversibility of the transferred energy in an ideal gas
in a model container depends on the degree of nonintegrability of
the single particle
Hamiltonian and is thus fully reversible for an integrable system. This
feature is illustrated in Fig.\ \ref{fig3} where the relative transferred
energy is shown over a full cycle of deformation for a cavity undergoing
harmonic spheroidal deformation. The surface of the spheroidal
cavity considered
is defined by

\begin{equation}
\frac{x^2}{a^2(t)} + \frac{y^2}{a^2(t)} + \frac{z^2}{c^2(t)}
=1,
\label{seqn11}
\end{equation}

\noindent
where

\begin{equation}
c(t)=R_0 (1 + \alpha (t)),
\label{seqn12}
\end{equation}

\noindent
and

\begin{equation}
a(t)=R_0 (1 + \alpha (t))^{-\frac{1}{2}},
\label{seqn13}
\end{equation}

\noindent
such that it undergoes volume conserving oscillations and particles
inside such a cavity constitute an integrable system. The deformation
will be given as $\alpha(t)= \alpha_0 - \alpha_0 \cos \omega t $,
where we have used $\alpha_0=1.0$ in our calculation.
We find in
Fig.\ \ref{fig3} that the transferred energy is fully reversible
and is independent of the speed of deformation.

For mixed (partially chaotic and partially regular) systems such as
cavities undergoing quadrupole deformations, a part of the transferred
energy is reversible while the rest is irreversible or dissipative
in nature. The chaos weighted wall formula (Eq.(\ref{seqn2}))
describes this irreversible part of the energy transfer. We shall
therefore proceed to obtain the dissipative part from the total
transferred energy as follows.

Decomposing the transferred energy $E_{trans}$ into its reversible
$E_{rev}$, and the dissipative $E_{diss}$, components, we write

\begin{equation}
E_{trans}(\eta,\alpha)=E_{rev}(\alpha) + E_{diss}(\eta, \alpha).
\label{seqn14}
\end{equation}

The incremental energy dissipation due to an incremental adiabaticity
index will therefore be given as,

\begin{equation}
\Delta E_{diss}(\Delta \eta,\alpha)= \Delta E_{trans}(\Delta \eta, \alpha),
\label{seqn15}
\end{equation}

\noindent
where

\begin{equation}
\Delta E_{diss}(\Delta \eta,\alpha)
= E_{diss}(\eta + \Delta \eta, \alpha)- E_{diss}(\eta , \alpha),
\label{seqn16}
\end{equation}

\noindent
and

\begin{equation}
\Delta E_{trans}(\Delta \eta,\alpha)
= E_{trans}(\eta + \Delta \eta, \alpha)- E_{trans}(\eta , \alpha).
\label{seqn17}
\end{equation}

We shall obtain $\Delta E_{trans}(\Delta \eta, \alpha)$ using
Eq.(\ref{seqn17}) by first evaluating $E_{trans}(\eta +
\Delta \eta, \alpha)$ and $ E_{trans}(\eta , \alpha)$ from classical
dynamical calculations. According to Eq.(\ref{seqn15}), this quantity
will ba same as $\Delta E_{diss}(\Delta \eta,\alpha)$  which
shall be compared with the chaos weighted wall formula prediction
in the next section.

We should remark here that one can ideally obtain the full dissipated
energy $E_{diss}(\eta, \alpha)$ from Eq.(\ref{seqn14})
by first calculating the reversible part as

\begin{equation}
E_{rev}(\alpha)
= \lim_{\eta \rightarrow 0} E_{trans}(\eta , \alpha).
\label{seqn18}
\end{equation}

This scheme is difficult to implement numerically though.
As the motion becomes slower (small $\eta$ ), computation time
increases prohibatively and numerical instability sets in.
Further, it may be recalled that
an interesting observation was made in Ref.\cite{BJS}
where it was shown that dynamical correlation between successive
collisions of a particle with the cavity wall can substantially enhance
the energy transfer rate beyond the wall formula limit for very slow
motions of the wall. Thus we shall not obtain the desired reversible
energy transfer from calculations at very slow
wall motions.
We have therefore
used finite values of $\eta$ (within adiabatic limit) in the
above incremental method for our purpose.

\section{RESULTS}

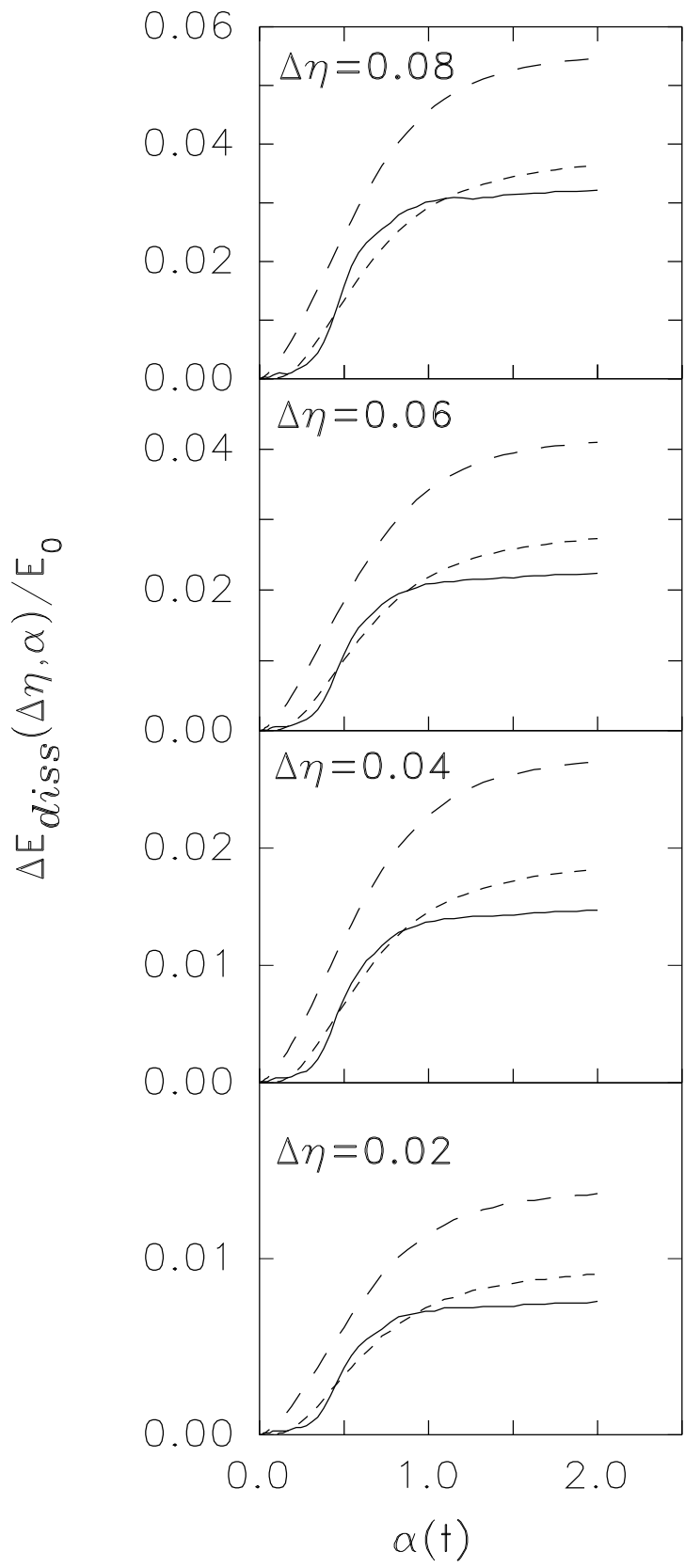
\begin{figure}
\caption
{Growth of incremental dissipated energy relative to $E_0$ for
different values of $\Delta \eta$. The solid lines are from
dynamical calculation.
The long dashed and short dashed lines are the original
wall formula (WF) and chaos weighted wall formula (CWWF) predictions
respectively.}
\label{fig4}
\end{figure}

We shall first consider a cavity which starts deforming from a
spherical shape and reaches a maximum deformation of $\alpha = 2.0$.
The relative energy transferred to the gas for this system at different
values of $\eta$ is given in Fig.\ \ref{fig2}. Following the procedure
outlined in the preceeding section, the incremental energy dissipation
for various values of $\Delta \eta$ are then obtained. This energy
dissipation  is  also calculated from both the chaos weighted
wall formula and the original wall formula using Eq.(\ref{seqn10}).
Fig.\ \ref{fig4} shows
the comparison. It is observed that the chaos weighted wall formula
prediction for energy dissipation is very close to that from dynamical
calculation. On the other hand, the original wall formula overestimates
the energy dissipation by more than $50 \%$.

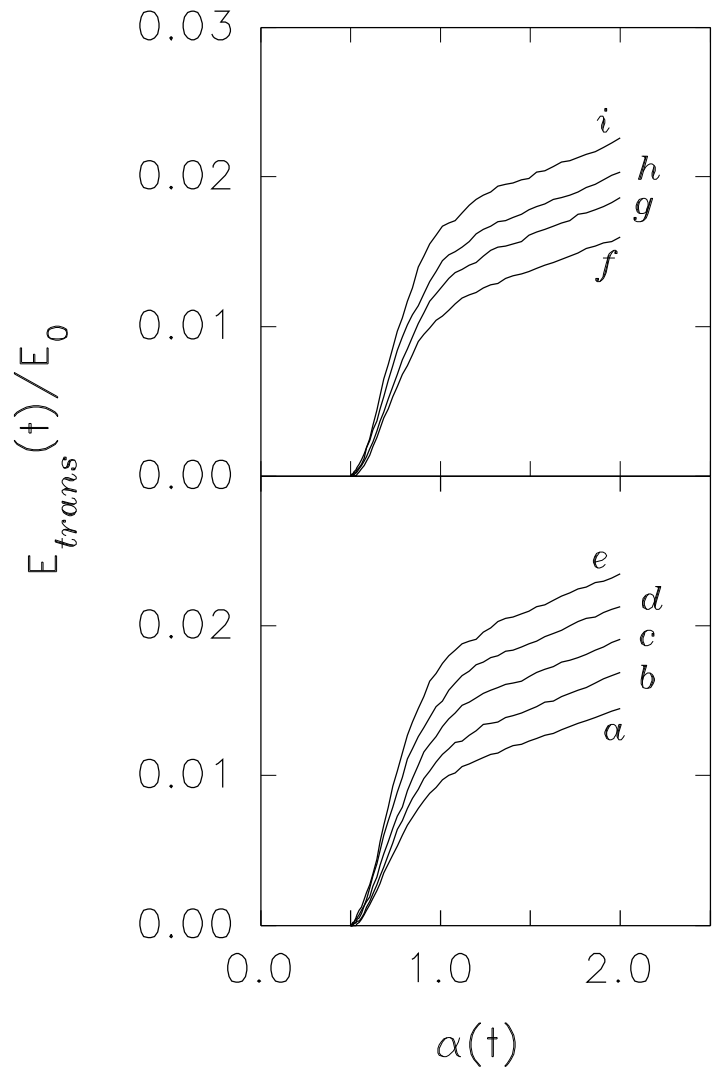
\begin{figure}
\caption
{Same as Fig.2 but for the cavity motion defined by
$\alpha_0 = 2.0$ and $\alpha_2 =1.5$.}
\label{fig5}
\end{figure}

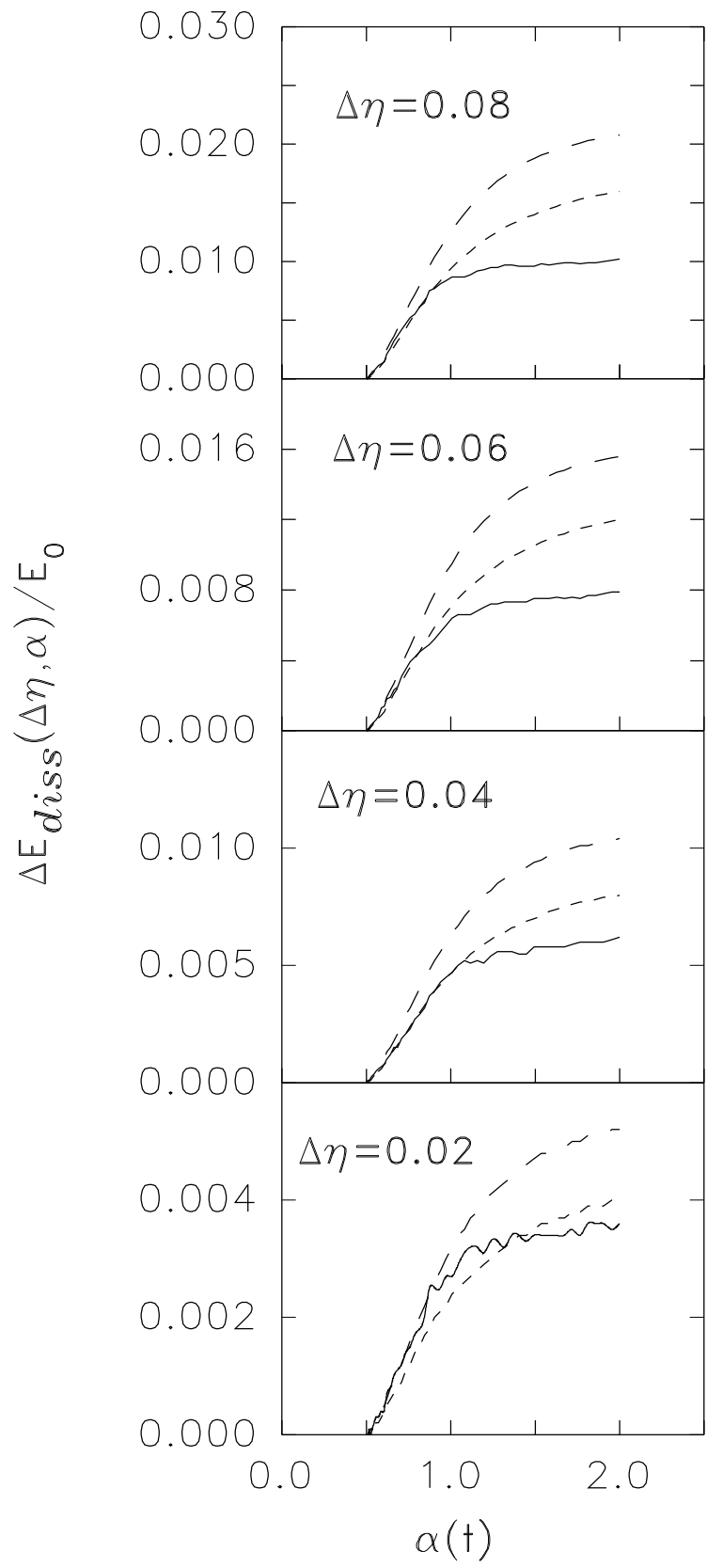
\begin{figure}
\caption
{Same as Fig.4 but for the cavity motion defined by
$\alpha_0 = 2.0$ and $\alpha_2 =1.5$.}
\label{fig6}
\end{figure}

We shall next consider a cavity
for which we set $\alpha_0 = 2.0$ and $\alpha_2 =1.5$.
This gives rise to an initial deformation of $\alpha = 0.5$ at $t=0$.
The results of dynamical calculations for energy transfer are
shown in Fig.\ \ref{fig5}. The incremental energy dissipations are
subsequently obtained  and the comparison with predictions
of the chaos weighted wall formula and the original wall formula
is shown in Fig.\ \ref{fig6}. The chaos weighted wall formula
prediction remains reasonably close to the dynamically calculated
values for lower values of $\Delta \eta$ (0.02 and 0.04). For
higher values of $\Delta \eta$ (0.06 and 0.08), the agreement becomes
somewhat poorer at larger deformations. Nevertheless, the chaos
weighted wall formula provides a much better description of the
dynamical results compared to the original wall formula for all the
cases considered here.

We shall now calculate the reduced friction coefficient which corresponds
to the dissipation according to the chaos weighted wall formula. The
reduced friction coefficient $\beta$ is defined \cite{FROB2} as
$\beta = \gamma / M$ where

\begin{equation}
\dot{E}=\gamma \dot{q}^2 R_0^2.
\label{seqn19}
\end{equation}

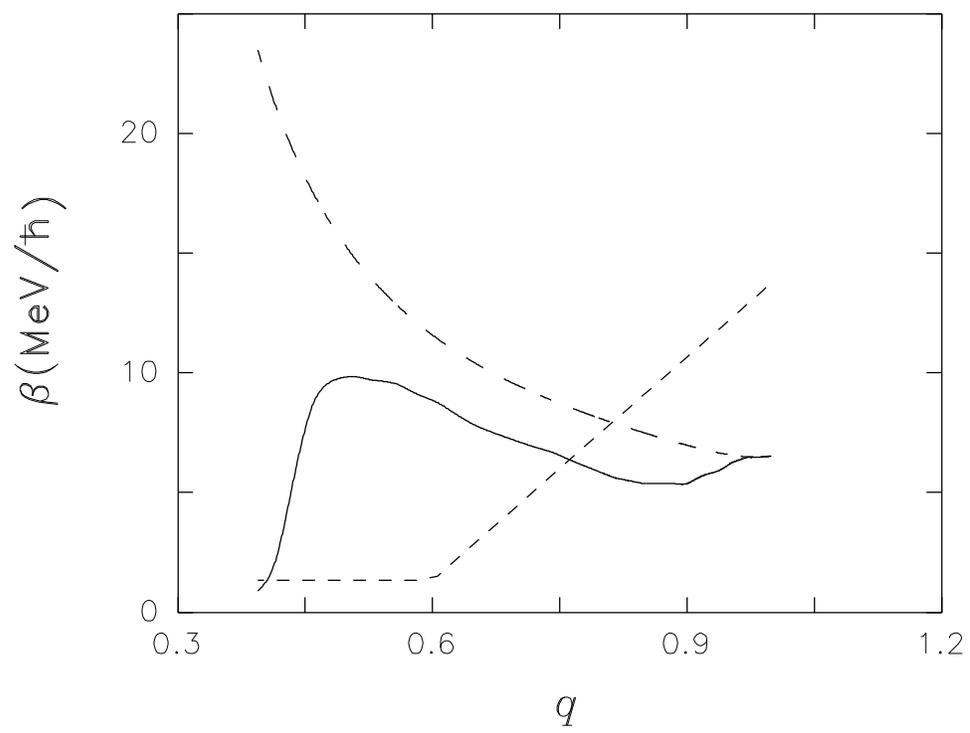
\begin{figure}
\caption
{Reduced friction coefficients from chaos weighted wall formula
(solid line), original wall formula (long dashed line) and
from Ref.[7] (short dashed line).}
\label{fig7}
\end{figure}

Here $q$ is half the distance between the centres of mass of the
equal fragments on each side of the neck
divided by $R_0$, $M$ is the total mass of the gas and
$\gamma$ is the friction coefficient.
The value of $q$ varies from $q=0.375$ for spherical configuration
to $q=1.0$ for scission.
The friction coefficient $\gamma$
and hence the reduced friction coefficient $\beta$ can be evaluated
from Eq.(\ref{seqn10}) for both the chaos weighted wall formula
and the original wall formula ($\mu=1$). The reduced friction coefficients
thus obtained are plotted in
Fig.\ \ref{fig7}. The phenomenological
reduced friction coefficient introduced by Fr\"{o}brich \cite{FROB1,FROB2}
is also shown in this figure.

We note in Fig.\ \ref{fig7} that the friction strength is strongly
suppressed in chaos weighted wall formula for spherical or near spherical
configurations. This reduction in the
dissipative force at the early stages of nuclear fission is an
important consequence of the chaos weighted wall formula.
It is of interest to note that a reduction of the wall plus window formula of
one body dissipation was also suggested \cite{NIX1,NIX2,NIX3} earlier.
Further, such a reduction
was phenomenologically found \cite{FROB2}
to be essential  in order to reproduce
the excitation functions
of both the prescission neutron multiplicities and the fission probabilities.
The subsequent increase of the reduced friction coefficient (due to
chaos weighted wall formula) at
larger values of $q$  essentially reflects the increase of chaoticity
with deformation. Since chaoticity depends sensitively on the
nonintegrability of the system, the reduced friction coefficient at
large $q$ in turn depends on the details of the shape considered.
Thus the comparison of our present
calculation for a model shape evolution
with the phenomenological friction is a qualitative one, both
showing an increase of the reduced friction at higher deformations
starting with very small values at the spherical configuration. We further
note that the phenomenological friction near scission is much stronger
than the original wall or its present chaos weighted version (both merge near
scission). This can possibly be understood \cite{ABE} if other effects such
as the exchange mechanism (''window'' formula) and temperature dependence
of nuclear friction are considered.

\section{ SUMMARY AND CONCLUSIONS}

In the preceding sections,
we have applied the chaos weighted wall formula to an ideal gas in
a cavity undergoing fissionlike shape evolution in order to test
the validity of the former for large amplitude motions. To this end, we have
dynamically calculated the energy transferred
 from the wall to the gas and compared its irreversible part  with the
prediction of the chaos weighted  wall formula. In order to simulate the
macroscopic dynamics of symmetric fission in our model, we have chosen
a cavity  initially at rest with an initial deformation
which is then progressivly deformed till the neck radius vanishes.
In all the cases considered above, we find a fair agreement between the
results from dynamical calculations and the predictions of the
chaos weighted wall formula. The original wall formula is found to
considerably overestimate the energy transfer.

We have also compared the reduced friction coefficient extracted from
the chaos weighted wall formula with that prescribed \cite{FROB2}
phenomenologically in order to reproduce the excitation functions
of both the precission neutron multiplicities and the fission probabilities.
Both the friction coefficients are found to be qualitatively similar.
An important feature
of the
 phenomenological friction is its strong suppression
for compact shapes which is also present in the chaos weighted wall
formula.

We can therefore conclude that the chaos weighted wall formula
provides a fairly accurate description of one body dissipation
in dynamical systems similar to fissioning nuclei.
Further, the qualitative similarity between the phenomenological friction
in nuclear fission and the chaos weighted wall formula
provides evidence for  one body nature of the dissipative force
acting in a fissioning nucleus.

\section*{ACKNOWLEDGEMENTS}

The  authors  are grateful to Professor Jan Blocki for sending us
the computer codes for classical  trajectory  calculations  which
have  been  used extensively in this work.

\appendix
\section{}
    We shall briefly outline here the essential steps to arrive at
the wall formula dissipation rate for cavities with large quadrupole
deformations which has been used in Eq.(\ref{seqn5}).

The wall formula is given as,
\begin{equation}
\dot{E}=\rho \bar{v} \int \dot{n}^2 d\sigma,
\label{seqn-owall}
\end{equation}

\noindent
where  $\dot{n}=$ normal component of the velocity of the surface
element $d\sigma$ at a point $\vec{r}=r(\theta,\phi)$.
The wall of an azimuthally symmetric quadrupole cavity is defined by,

\begin{equation}
r(\theta ,t)=\frac{R_0}{\lambda(t)}[1+
                     \alpha(t) P_2 (\cos\theta)].
\label{seqn-surf}
\end{equation}

Taking the surface velocity component along
the normal,

\begin{equation}
\dot{n}=\dot{r}\cos\theta_n,
\end{equation}

\noindent
where $\theta_n$ is the angle between $\vec{r}$ and $\vec{n}$ and
is given by,

$$\cos\theta_n  =  r \frac{d\theta}{ds}  =
\frac{r}{[r^2+(\frac{dr}{d\theta})^2]^{1/2}}, $$

\noindent
and where we have used,

$$ ds
     =d\theta \big[r^2+\big(\frac{dr}{d\theta}\big)^2\big]^{1/2}. $$

Hence
\begin{equation}
\dot{E}
 =  2\pi\rho\bar{v}\int \frac{\dot{r}^2 r^3}
           {[r^2+(\frac{dr}{d\theta})^2]^{1/2}  }\sin\theta d\theta
\label{seqn-bigw}
\end{equation}

Now from Eq.(\ref{seqn-surf}) we get,

\begin{equation}
\dot{r}=\frac{R_0}{\lambda}[\dot{\alpha}P_2 -
\frac{\dot{\lambda}}{\lambda}(1 +\alpha P_2) ],
\label{seqn-rdot}
\end{equation}

\noindent
and

\begin{equation}
\frac{dr}{d\theta}=-\frac{R_0}{\lambda}
          \alpha P^{\prime}_{2} (z)\sin\theta ,
\label{seqn-drdth}
\end{equation}

\noindent
where  $P^{\prime}_{2}(z)=dP_2(z)/dz$ and $z=\cos\theta$. Using the above
in Eq.(\ref{seqn-bigw}), we obtain

\begin{mathletters}
\begin{equation}
\dot{E}=\frac{2\pi\rho\bar{v}R_0^4 \dot{\alpha}^2}{\lambda^4}
    \int_{-1}^{+1} F(z)dz  \\
\end{equation}
\text{where}
\begin{equation}
F(z)=\frac{ [P_2(z)-\frac{1}{\lambda}(\frac{d\lambda}{d\alpha})
\big(1+\alpha P_2(z)\big)]^2 \big(1+\alpha P_2(z)\big)^3 }
{[\big(1+\alpha P_2(z)\big)^2 +\alpha^2 P^{\prime^2}_{2}(z)
          (1-z^2)]^{1/2}}.
\end{equation}
\label{seqn-lrd}
\end{mathletters}

The  above equations  give  the  rate of energy transfer for
large quadrupole deformations.

\end{document}